\documentclass[a4paper]{article}

\usepackage{xspace}
\usepackage{INTERSPEECH2021}
\usepackage{multirow}
\usepackage{cite}
\usepackage{hyperref}
\usepackage{color}
\usepackage{subfigure}
\usepackage{algorithm}
\usepackage{algorithmic}
\title{Cross-domain Speech Recognition with Unsupervised Character-level Distribution Matching}
\name{Wenxin Hou$^1$, Jindong Wang$^2$, Xu Tan$^2$, Tao Qin$^2$, Takahiro Shinozaki$^1$\thanks{Work done when the first author was in internship at MSRA. Correspondence to: Jindong Wang.}}
\address{
  $^1$Tokyo Institute of Technology\\
  $^2$Microsoft Research Asia}
\email{hou.w.aa@m.titech.ac.jp, \{jindong.wang, xuta, taoqin\}@microsoft.com, shinot@ict.e.titech.ac.jp}

\newcommand{\method}{CMatch\xspace}

\newcommand{\argmax}{\arg \max}

\begin{document}
\maketitle
\begin{abstract}
End-to-end automatic speech recognition (ASR) can achieve promising performance with large-scale training data. However, it is known that domain mismatch between training and testing data often leads to a degradation of recognition accuracy. In this work, we focus on the unsupervised domain adaptation for ASR and propose CMatch, a \underline{C}haracter-level distribution matching method to perform fine-grained adaptation between each character in two domains. First, to obtain labels for the features belonging to each character, we achieve frame-level label assignment using the Connectionist Temporal Classification (CTC) pseudo labels. Then, we match the character-level distributions using Maximum Mean Discrepancy. We train our algorithm using the self-training technique. Experiments on the Libri-Adapt dataset show that our proposed approach achieves 14.39\% and 16.50\% relative Word Error Rate (WER) reduction on both cross-device and cross-environment ASR. We also comprehensively analyze the different strategies for frame-level label assignment and Transformer adaptations.
\end{abstract}

\section{Introduction}

With the recent success of deep learning and high-quality, large-scale transcribed corpora, the performance of end-to-end (E2E) automatic speech recognition (ASR) has been significantly improved.
Existing approaches are built on the i.i.d. condition that training and testing data are from the same distribution, e.g., from the same recording device or environment.
However, this assumption does not always hold in reality when the training and test distributions are different.
For instance, a well-trained ASR model based on PlayStation Eye recordings is likely to have deteriorated performance when it is used to recognize speech from the Matrix recordings.
It is expensive and time-consuming to collecting labeled speech data from massive domains (distributions).
In this paper, we tackle a more challenging scenario where there are no labeled samples available in the test data (i.e., target domain).
Thus, our goal is to perform unsupervised domain adaptation (UDA)~\cite{pan2009survey,patel2015visual} to improve the cross-domain ASR performance on the unlabeled target domain using the well-labeled source data.

UDA for ASR has been studied in the existing literature.
Liang et al.~\cite{liang2018learning} proposed to combine data augmentation with representation matching to force the model to learn noise-invariant representations between clean speech and their augmented noisy counterparts.
In~\cite{khurana2020unsupervised}, the authors were able to recover 60\% to 80\% of the word error rates (WER) on the target domain by introducing a pseudo-label filtering approach based on the model's uncertainty using dropout for ASR UDA.
Recent work \cite{sun2018domain,duan2020unsupervised} used the domain-adversarial training~\cite{ganin2016domain} for speech recognition where they adversarially trained domain discriminators to distinguish the source and target samples.

\begin{figure}[t!]
    \centering
    \subfigure[Before \method]{
    \includegraphics[width=.22\textwidth]{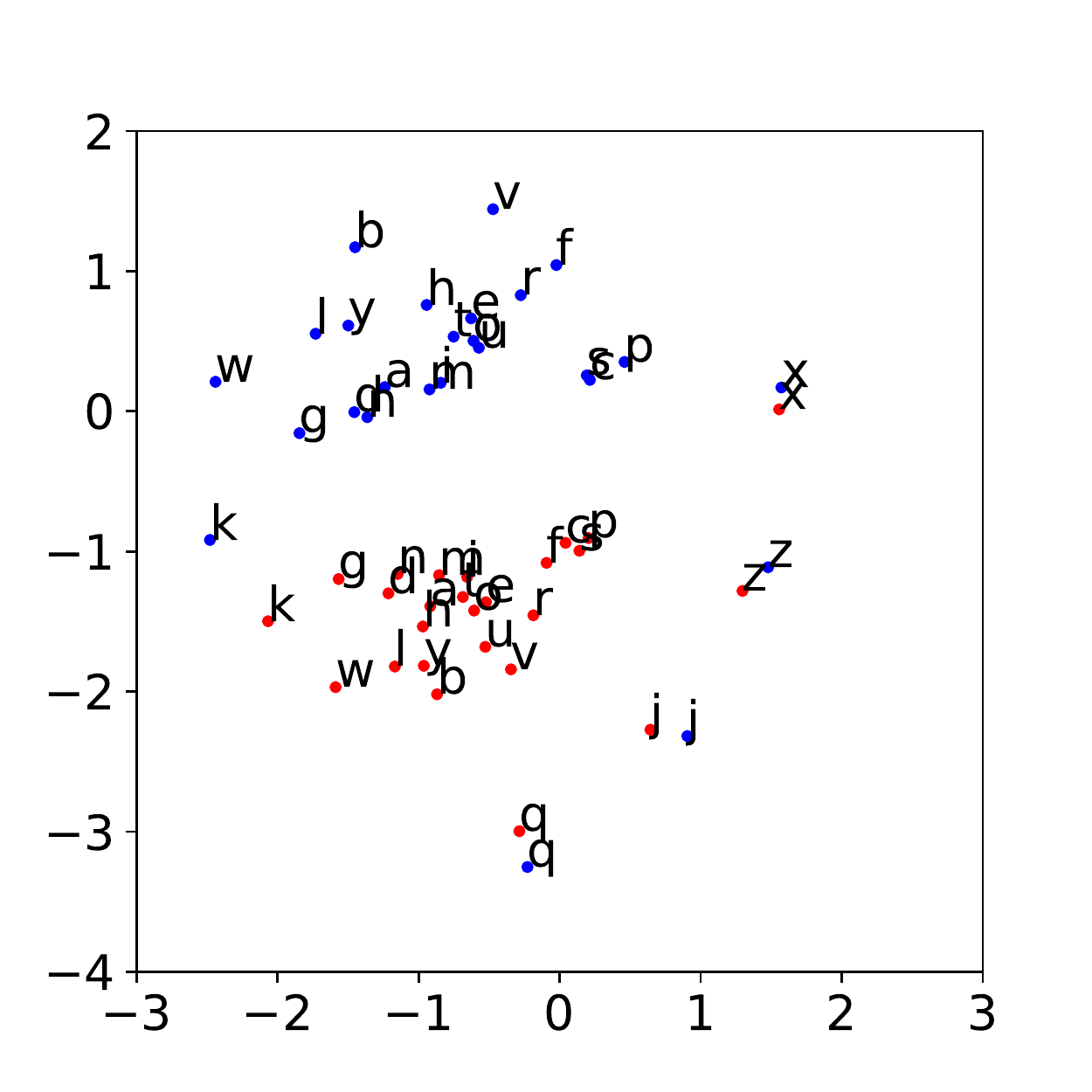}
    \label{fig-motiv-before}
    }
    \subfigure[After \method]{
    \includegraphics[width=.22\textwidth]{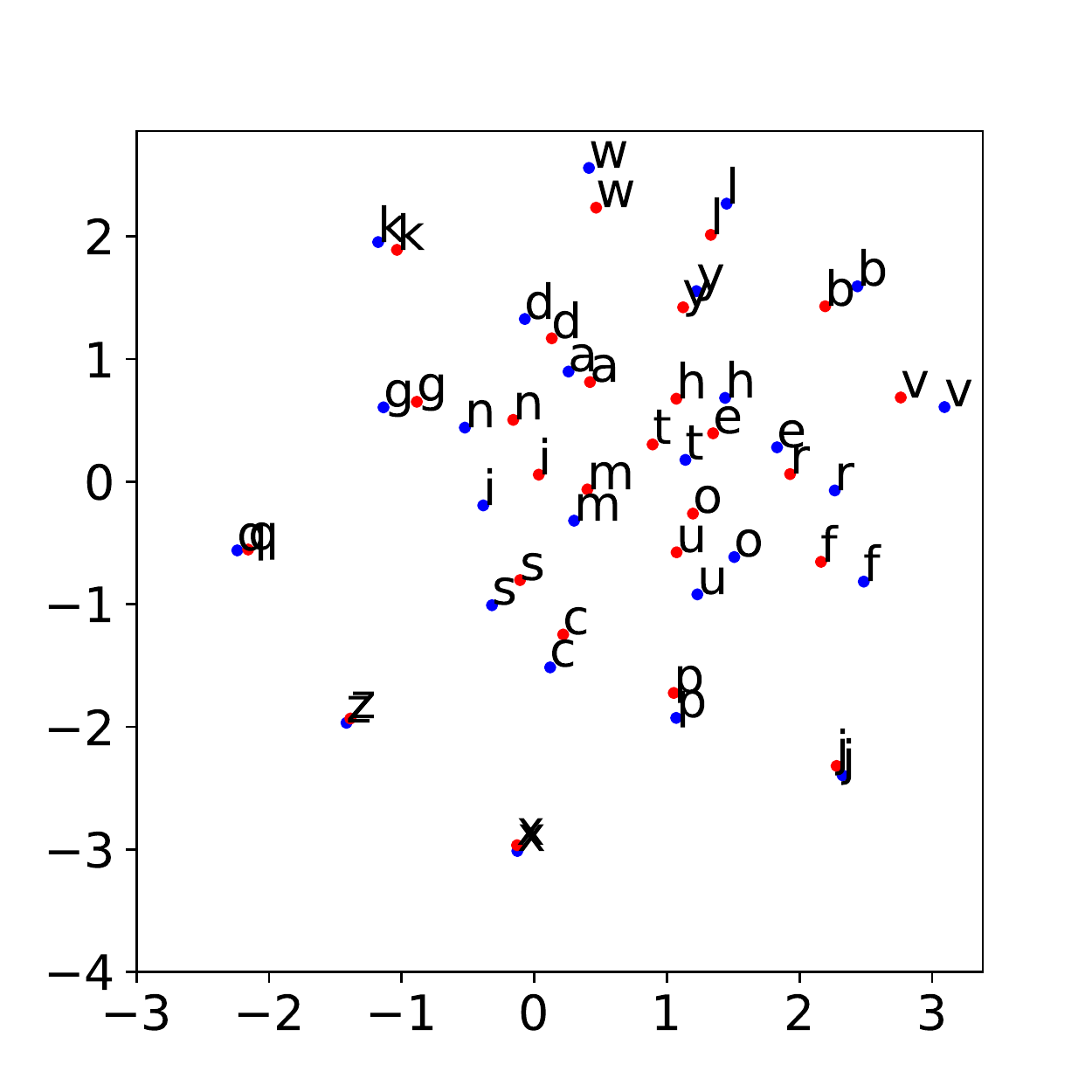}
    \label{fig-motiv-after}
    }
    \vspace{-.1in}
    \caption{Feature centers of all characters in two domains are closer after \method. Red and blue dots denote the source and target domains, respectively. Some obvious bad cases in (a): \{a, b, c, g, h, l, r, v, w, y\} are aligned well in (b).}
    \label{fig-motiv}
    \vspace{-0.4cm}
\end{figure}

However, existing methods typically treated a domain as one distribution when matching distributions.
They may ignore the fine-grained knowledge in cross-domain ASR tasks that can be utilized for distribution matching.
These fine-grained knowledge are important to preserve the detailed distribution property of different domains, such as the characters, phoneme, and word.
Ignoring these information is likely to result in unsatisfying results.
This is also validated in \cite{zhu2020deep} where the images aligned in subdomains (i.e., domains split by class labels) can generally achieve better adaptation performance than traditional methods that align the whole domains.

In this paper, we propose \method, an algorithm to match the \underline{C}haracter-level distributions for cross-domain ASR.
Compared to other types such as word-level that are sparse and could lead to unstable performance, the character-level distribution is more fine-grained and easier to compute in E2E ASR models.
Thus, it can preserve more fine-grained knowledge for each character than existing methods that used one distribution for all characters. \figurename~\ref{fig-motiv} shows that the distances of the same characters from two domains are smaller after applying our algorithm.

To enable character-level distribution matching, \method consists of two steps.
First, we achieve frame-level label assignment to acquire the labels for each encoded frame to compute the conditional distribution.
This is done by using the Connectionist Temporal Classification (CTC)~\cite{graves2006connectionist} pseudo labels. 
Second, \method can reduce the conditional distributions between each cross-domain characters using the well-defined Maximum Mean Discrepancy (MMD)~\cite{gretton2012kernel}.
The above two steps are jointly optimized in a self-training framework~\cite{masumura2020sequence} in Speech-Transformer~\cite{dong2018speech}.
Experiments on the Libri-Adapt dataset~\cite{mathur2020libri} show that our proposed approach achieves 14.39\% and 16.50\% relative Word Error Rate (WER) reduction on both cross-device and cross-environment ASR.

\section{Preliminary}
\subsection{CTC-Attention Transformer ASR Model}
In this work, we built our E2E ASR model based on the Speech-Transformer~\cite{dong2018speech} with joint CTC-attention structure~\cite{kim2017joint} which has been successfully applied in various ASR tasks~\cite{karita2019comparative, Hou2020, hou2021meta, hou2021exploiting}. 

As model inputs, the acoustic features are 83-dimensional filter banks with pitch features computed with 10 ms frame shift and 25 ms frame length. The acoustic features are further subsampled by a factor of 4 by 2 convolution layers with kernel size 3 and stride 2. The resulted features have a receptive field of 100 milliseconds for each frame. Then the following encoder layers process the subsampled features by
self-attention and feed-forward as illustrated in~\cite{vaswani2017attention}. Apart from self-attention and feed-forward, the decoder layers further accept the encoder outputs and perform cross-attention.

For the CTC-attention structure, an auxiliary CTC task~\cite{graves2006connectionist} is introduced for encoder outputs in order to encourage the monotonic alignment and accelerate the convergence speed. In training, a multi-task learning scheme is adopted consisting of two losses: the attention loss $\mathcal{L}_{\text{ATT}}$ and the CTC loss $\mathcal{L}_{\text{CTC}}$:
\begin{equation}
\label{eq:asrloss}
    \mathcal{L}_{\text{ASR}} = (1 - \lambda) \mathcal{L}_{\text{ATT}} + \lambda \mathcal{L}_{\text{CTC}},
\end{equation}
where $\lambda$ denotes the weight of the CTC module.
Similarly, during decoding, the beam search decoding results of the Transformer decoder are re-scored by the CTC module:
\begin{equation}
    \hat{Y} = \arg \max_{Y\in \mathcal{Y}} (1 - \lambda) \log P_{\text{ATT}} (Y|X) + \lambda \log P_{\text{CTC}}(Y|X),
\end{equation}
where $X$ denotes the acoustic features, $\mathcal{Y}$ represents a set of the decoding hypotheses.



\section{Our Method: \method}
\label{sec-method}

The goal of cross-domain ASR is to predict the transcript for the target speech given only paired source domain data $(X_S, Y_S)$ and the speech-only target domain data $X_T$. 
Our key idea is to match the character-level distributions, which can be computed via the conditional distributions $P(Y|X)$ (i.e., features belonging to each character in ASR).
To this end, there are two technical challenges ahead.
First, how to acquire the conditional distributions for the speech features, i.e., obtain the features for each character is difficult in this sequence-to-sequence framework since we do not have one-to-one mapping for the encoded speech features and labels.
Second, how to reduce the distribution divergence between these conditional distributions is another challenge.



\subsection{Frame-level Label Assignment}
\label{ssec-align}
Computing $P(Y|X)$ requires obtaining the labels for each input, which refers to the frame-level labels in ASR as its inputs are frames.
The feature of an input $x$ extracted by Transformer encoder $f$ is $f(x) \in  \mathbb{R}^{N \times D}$, where $N$ denotes numbers of frames and $D$ denotes feature dimension.
The Transformer decoder will output $M$ labels using CTC, which is not mapped to the $N$ frames, i.e., $N \ne M$.
Thus, it is challenging to acquire frame-level labels $y$ for computing conditional distributions $P(Y|X)$.
Note that this challenge exists for both the source and target domains.
This is also significantly different from image classification where we can easily get the labels for each sample~\cite{pan2019transferrable,zhu2020deep}.

\begin{figure}[t]
    \centering
    \includegraphics[width=.48\textwidth]{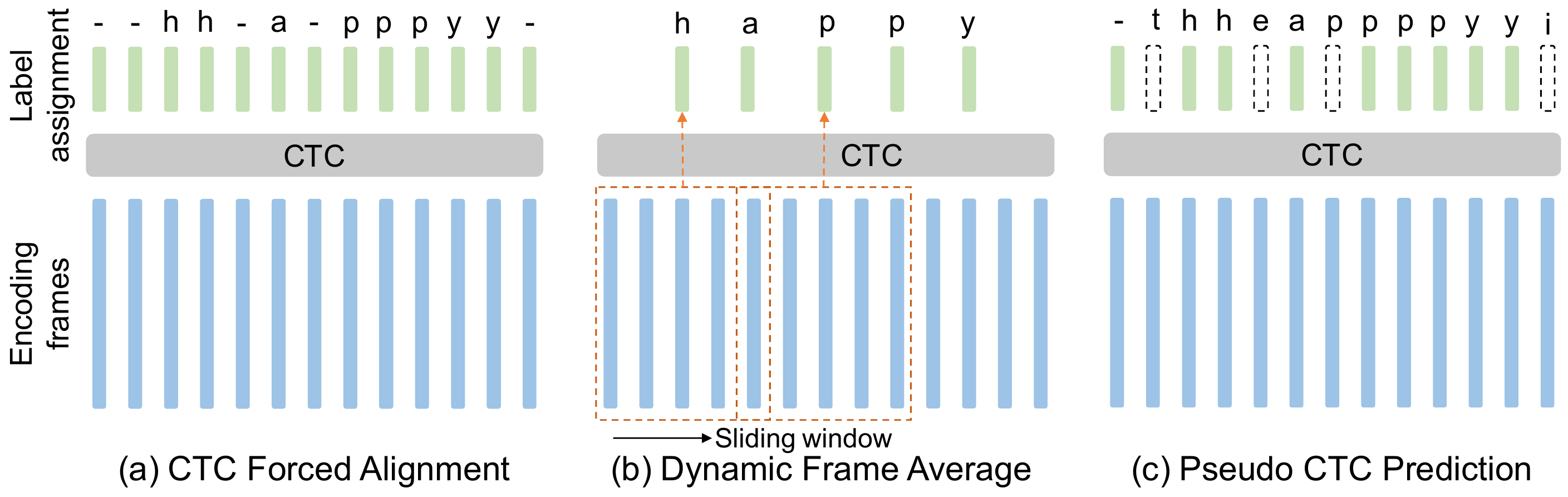}
    \caption{Three strategies for frame-level label assignment. Symbol ``-'' represents {\ttfamily<blank>}. Dotted labels are filtered out based on their predicted confidence scores.}
    \vspace{-.2in}
    \label{fig-align}
\end{figure}

It is effective to use CTC forced alignment~\cite{sak2015learning, kurzinger2020ctc} to take the labels from the most probable path selected by CTC as the frame-level assignment (\figurename~\ref{fig-align}(a)).
However, this process is computationally expensive.
It is also feasible to use a dynamic window average (\figurename~\ref{fig-align}(b)) strategy that assigns labels for each frame by averaging, which can only work in a strict condition that the character output is a uniform distribution.
In our work, to obtain the frame-level labels, we propose to use the CTC pseudo labels (\figurename~\ref{fig-align}(c)) for both efficiency and correctness guarantee.
We assume that an ideal CTC model should predict the label assignment with high accuracy and propose to directly utilize the CTC predictions since the CTC module naturally predicts labels frame by frame including {\ttfamily<blank>} symbol as the null label.
More formally, the pseudo label for the $n$-th frame $X_n$ can be obtained by:
\begin{equation}
\label{eq-ctcalign}
    \hat{Y}_n = \argmax_{Y_n} P_\text{CTC}(Y_n|X_n), \quad 1 \le n \le N.
\end{equation}

We further filter out the CTC predictions with a threshold based on their softmax scores to improve the accuracy. Only the labels with a softmax score of over 0.9 are used.


\subsection{Distribution Matching}
\label{ssec-mmd}
For distribution matching, we adopt the popular 
Maximum Mean Discrepancy (MMD)~\cite{borgwardt2006integrating} for distribution measure.
MMD is an efficient non-parametric criterion to empirically evaluate the distribution divergence between two domains. MMD has been widely applied in the fields of computer vision, natural language processing, etc.~\cite{pan2010domain, tzeng2014deep, zhang2017joint}, and its effectiveness has been theoretically proved~\cite{gretton2012kernel}.
Given $X_S \sim P, X_T \sim Q$, the MMD distance between two distributions $P$ and $Q$ can be formulated as follows:
\begin{equation}
    \text{MMD}(\mathcal{H}_k, P, Q) = \sup_{||\phi||_{\mathcal{H}_k \leq 1}} \mathbb{E}_{X_S \sim P} \phi(X_S) -  \mathbb{E}_{X_T \sim Q} \phi(X_T),
\end{equation}
where $\mathcal{H}_k$ is the Reproduced kernel Hilbert space with Mercer kernel $k(\cdot, \cdot)$, $||\phi||_{\mathcal{H}_k} \leq 1$ denotes its unit norm ball and $\phi(\cdot)$ is the feature map.

A biased empirical estimate of MMD can be obtained by computing the empirical expectations on the samples:
\begin{equation}
\begin{split}
    &\text{MMD}(\mathcal{H}_k, X_S, X_T) =
    \\ 
    &\sup_{||\phi||_{\mathcal{H}_k \leq 1}} \left( \frac{1}{|X_S|}\sum_{x_s\in X_S} \phi(x_s)- \frac{1}{|X_T|}\sum_{x_t \in X_T} \phi(x_t) \right).
\end{split}
\end{equation}

The character-level distribution matching is computed as:
\begin{equation}
    \mathcal{L}_{\text{cmatch}} = \frac{1}{|\mathcal{C}|} \sum_{c \in \mathcal{C}} \text{MMD}(\mathcal{H}_k, X_S^c, X_T^c),
\end{equation}
where $X_S^c$, $X_T^c$ denotes the encoder features of the source and target samples of the class $c$, $\mathcal{C}$ denotes the character set. Note that we use CTC pseudo labels for both of the source and target domains, instead of using ground-truth labels. For real implementation, we input the features extracted by Transformer encoder to MMD instead of the raw inputs $X$ for computation.

\subsection{Overall Loss}
Finally, we apply self-training by utilizing the pseudo labeled target-domain data $(X_T, \hat{Y}_T)$ with the pseudo ASR loss $\mathcal{L}_\text{ASR}^\text{tgt}$ to assist the distribution matching. Given the speech-only target domain data $X_T$ and the ASR model pre-trained on source domain, we can obtain the beam search decoding results as pseudo labels $\hat{Y}_T$ for the target speech data $X_T$. Denote $\gamma$ the tradeoff hyperparameter for the distribution matching loss, the overall formulation of \method is:
\begin{equation}
\label{eq-all}
    \mathcal{L} = \frac{1}{2}\left(\mathcal{L}_\text{ASR}^\text{src} + \mathcal{L}_\text{ASR}^\text{tgt} \right) + \gamma \mathcal{L}_\text{cmatch}.
\end{equation}

To reduce the noise in pseudo labels, we filter out the last 30\% utterances based on their beam search confidence scores.
The complete algorithm for \method is illustrated in Algo.~\ref{algo}.
Note that Step 2 in the algorithm (i.e., self-training) can be iterated while we found it is enough to do it once in experiments to achieve satisfying results.

\begin{algorithm}[t!]
	\caption{Learning algorithm of \method}
	\label{algo}
	\textbf{Input}: Source domain $(X_S, Y_S)$, target domain $X_T$.
	
	\begin{algorithmic}[1] 
		\STATE Train network $M_S$ on source domain $(X_S, Y_S)$.
		\STATE Obtain pseudo label $\hat{Y}_T$ with $M_S$.
		\WHILE{not done}
		\STATE Obtain the frame-level labels using Eq.~\eqref{eq-ctcalign}.
		\STATE Joint optimization using Eq.~\eqref{eq-all}.
		\ENDWHILE
		\STATE \textbf{return} Adapted model $M_{S \to T}$ and target transcripts.
	\end{algorithmic}
\end{algorithm}


\section{Experimental Setup}
\subsection{Data Set}
We employ the Libri-Adapt dataset~\cite{mathur2020libri} for the experiments. The Libri-Adapt dataset is designed for unsupervised domain adaptation tasks and built on top of the Librispeech-clean-100 corpus recorded using 6 microphones under 4 synthesized background noise conditions (Clean, Rain, Wind, Laughter) in 3 different accents (en-us, en-gb, and en-in). Since the cross-accent data has not been not fully released by the author, in this work, we use the US accent (en-us) as the main accent language and keep it unchanged to isolate the influence brought by accents.

We constructed two types of cross-domain ASR tasks: (1) cross-device ASR adaptation, i.e., the source and target devices are different; and (2) cross-environment ASR adaptation, i.e., the source and target environment are different.
Specifically, we select 3 devices for cross-device experiments, namely Matrix Voice (M), PlayStation Eye (P), and ReSpeaker (R), each two of them can form a pair of tasks, leading to 6 tasks in total.
For cross-environment ASR, we select 3 types of noise (Rain, Wind, and Laughter) for different target environments and the source environment remains clean, leading to 3 tasks.
In our experiments, we split the last 10\% utterances in the training set (source domain) for development. For each domain, the numbers of training/validation/test utterances are 25685/2854/2600 corresponding to hours of 93.77/10.71/5.60 hours, respectively.
Note that we did not use any of the labels on the target domain during training, and they are only used for evaluation.


\subsection{Baselines}
We consider the following approaches for comparison:
\begin{itemize}
    \item Source-only: directly apply the pre-trained model from the source domain to target domain without adaptation.
    \item MMD-ASR: this is used in recent literature for cross-domain speech recognition~\cite{shivakumar2020transfer}. We minimize the MMD distance on the averaged encoder outputs to encourage the domain-level feature alignment.
    \item Domain Adversarial Training (ADV): this is adopted in recent work~\cite{sun2018domain,duan2020unsupervised}. An adversarial domain classifier is trained on top of the averaged encoder outputs to correctly classify the feature domains, while the encoder is trained to maximize the domain classification loss. We also re-implemented this idea.
\end{itemize}

\subsection{Implementation Details}
We implemented our experiments based on the ESPnet codebase~\cite{Watanabe2018}.
The Speech-Transformer has 12 encoder layers and 6 decoder layers with 4 attention heads and an inner dimension of 2048.
We use 31 characters including 26 alphabets and 5 symbols {\ttfamily (<'>, <space>, <eos>, <unk>, <blank>)} as the modeling units. The weight of the CTC module is set to 0.3 for both training and decoding. We set $\gamma$ to 10.0 for our \method method tuned on the validation set. For the baseline MMD and ADV methods, adaptation loss weight $\gamma$ is set to 10.0 and 0.3 based on the validation set, respectively.
We set $\lambda=0.3$ following ESPnet~\cite{Watanabe2018}.
For both pseudo labeling and decoding, beam size of 10 is used.
We choose the primal kernel for the MMD kernel.
We adapt the final encoder layer of the Transformer, which is the same as MMD and ADV methods for fair study.
We train all the models for 100 epochs with batch size 64. Early stopping is adopted with patience 10 and 5 epochs during pre-training and adaptation, respectively. We optimize the models in the same way as~\cite{vaswani2017attention} with warmup steps 25000 and a maximum learning rate of 3.0.
Code for \method is at \url{https://github.com/jindongwang/transferlearning/tree/master/code/ASR/CMatch}.


\section{Results}

For comparison, we first show the word error rates (WER) of the in-domain ASR results in \tablename~\ref{tb-baseasr} where the training and test data are from the same domain.

\begin{table}[htbp]
    \centering
    \caption{Results (WER) on standard ASR in clean environment.}
    \vspace{-.1in}
    \begin{tabular}{cc}
    \toprule
        Domain &  WER \\
    \hline
        Matrix Voice (M) & 24.25 \\
        PlayStation Eye (P) & 20.07 \\
        ReSpeaker (R) & 23.78 \\
    \hline
        Average & 22.70 \\
    \bottomrule
    \end{tabular}
    \label{tb-baseasr}
\end{table}

\subsection{Cross-domain Adaptation Results}

\textbf{Device adaptation}.
We conduct cross-device ASR experiments in \tablename~\ref{tb-crossdevice}.
Compared to \tablename~\ref{tb-baseasr}, the baseline results show a severe performance degradation caused by distribution shift.
Compared to source-only method, both adversarial training and MMD-ASR approaches obtain improvements on the target domain with unlabeled target-domain speech only.
MMD surpasses ADV on most tasks, which demonstrates the effectiveness of the MMD.
Our \method achieves the lowest WERs on all tasks and obtains an average WER of 22.85\% with a relative reduction of $\mathbf{14.39\%}$ compared with the baseline.


\begin{table}[t]
\centering
\caption{WER on cross-device ASR in clean environment.}
\label{tb-crossdevice}
\begin{tabular}{ccccc}
\toprule
Task          & Source-only & MMD   & ADV   & \method \\
\hline
M $\to$ P     & 23.87       & 20.87 & 21.11 & \textbf{20.38}      \\
M $\to$ R     & 25.21       & 22.21 & 22.27 & \textbf{21.77}      \\
P $\to$ M     & 31.15       & 27.22 & 28.29 & \textbf{26.17}      \\
P $\to$ R     & 23.99       & 21.90  & 21.74 & \textbf{20.43}      \\
R $\to$ M     & 32.45       & 28.27 & 29.95 & \textbf{27.77}      \\
R $\to$ P     & 23.48       & 21.09 & 21.23 & \textbf{20.58}      \\
\hline
Average & 26.69       & 23.59 & 24.10 & \textbf{22.85}    \\
\bottomrule
\end{tabular}
\end{table}

\vspace{0.1cm}
\noindent\textbf{Noise adaptation.}
We pick the ReSpeaker device for the noise adaptation experiments, whose standard WER in the clean environment is 23.78\%.
The results are presented in \tablename~\ref{tb-noise}.
We also observe similar performance of MMD and ADV.
Our \method achieves $\mathbf{16.50\%}$ relative WER reduction on average, indicating its effectiveness for cross-environment adaptation.
\method still shows the best performance on all the tasks. It is worth noting that when adapted to the wind environment, \method obtains very competitive results without the real labels.
 
\begin{table}[t!]
\caption{WER on cross-environment adaptation (source: Clean)}
\label{tb-noise}
\centering
\begin{tabular}{ccccc}
\toprule
Target & \multicolumn{1}{c}{Source-only} & MMD   & ADV   & \method \\
\hline
Rain          & 38.21                           & 33.61 & 34.65 & \textbf{32.90}  \\
Wind          & 29.70                            & 26.06 & 26.73 & \textbf{23.12}  \\
Laughter      & 33.36                           & 29.85 & 30.41 & \textbf{28.55}  \\
\hline
Average       & 33.76                           & 29.84 & 30.60 & \textbf{28.19}  \\
\bottomrule
\end{tabular}
\vspace{-.2in}
\end{table}

\vspace{-0.1cm}
\subsection{Ablation Study}

We conduct the ablation study in \tablename~\ref{tab:ablation}.
Results indicate that compared to baselines, our character-level matching loss and the self-training strategy can independently produce competitive performance.
Thus, they are critical in our algorithm.
We observe that self-training can generally perform well, indicating its effectiveness in obtaining labels, which is important in our unsupervised conditional distribution matching algorithm.
Besides, we also tested the combination of vanilla MMD and self-training and found its performance not comparable to our results.
By combining the advantage of fine-grained distribution matching and self-training, our \method obtains the best performance through fine-grained distribution matching.

\begin{table}[t]
    \centering
    \caption{Ablation study on the \method method.}
    \label{tab:ablation}
    \begin{tabular}{ccc}
    \toprule
               Variant    & Device & Noise \\
    \hline
        Source-only & 26.69 & 33.76 \\
         w/ self-training & 22.99 & 28.31 \\
         w/ distribution matching & 23.87 & 30.43 \\
         All & 22.85 & 28.19 \\
    \bottomrule
    \end{tabular}
\end{table}

\vspace{-0.1cm}
\subsection{Analyzing the Label Assignment in \method}
In this section, we compare three label assignment approaches as mentioned in Section~\ref{ssec-align}. The results in \tablename~\ref{tb-align} show that the CTC alignment method achieves the best results. However, we found it computationally expensive ($2\times$ slower than others since the forced alignment requires computing the forward-backward algorithm in each iteration).
As its complement, frame average also achieves good results, but it can generally works with uniform speaking speed.
Our pseudo CTC prediction has no assumption on the speaking speed while obtaining similar results and more efficient than CTC alignment.

\begin{table}[tb]
\caption{Comparison of different label assignment strategy.}
\label{tb-align}
\begin{tabular}{ccccc}
\toprule
Task & PseudoCTCPred & FrameAverage & CTCAlign \\
\hline
M $\to$ P       & 20.38        & 20.21                 & 20.23                 \\
M $\to$ R       & 21.77        & 21.80                 & 21.75                 \\
P $\to$ M       & 26.17        & 26.02                  & 25.84                 \\
P $\to$ R       & 20.43        & 20.36                 & 20.44                \\
R $\to$ M       & 27.77        & 27.94                 & 27.73                 \\
R $\to$ P       & 20.58        & 20.55                 & 20.52                \\
\hline
Average   & 22.85              & 22.81                 & 22.75               \\
\bottomrule
\end{tabular}
\vspace{-.2in}
\end{table}

\vspace{-0.1cm}
\subsection{Adapting Encoder and Decoder}

A fundamental step in cross-domain ASR is to determine the adaptation contribution of the encoder and decoder.
In our experiments, we adapt the last encoder layer.
It is intuitive and necessary to ask: Will adapting the encoder and decoder lead to better performance?
In this section, we empirically answer these questions via extensive experiments. \tablename~\ref{tb-dec-ablation} shows our results, where ``first'', ``last'', and ``all'' denote we adapt the first / last / all decoder layers, respectively.
It implies that in ASR, it is more necessary to adapt the encoder layers.
Since the decoders are not task-specific in our experiments (all standard English ASR), reducing their distribution gap may not be effective and could generally hurt the performance.
The conclusions are similar on cross-device experiments.
In addition, we did not get positive results by adapting more encoder layers. Therefore, adapting the last encoder layer is sufficient in our task.

\begin{table}[h]
\caption{\method with / without decoder adaptation. ``first'', ``last'' and ``all'' correspond to decoder layer(s) for adaptation.}
\centering
\label{tb-dec-ablation}
\begin{tabular}{ccccc}
\toprule
Target & w/o decoder    & first & last  & all   \\
\hline
Rain                   & 32.90 & 32.92 & 32.85 & 33.12 \\
Wind                   & 23.12 & 23.18 & 23.18 & 23.28 \\
Laughter               & 28.55 & 28.66 & 28.56 & 28.63 \\
\hline
Average                & 28.19          & 28.25 & 28.20 & 28.34 \\
\bottomrule
\end{tabular}
\vspace{-0.35cm}
\end{table}

\vspace{-0.2cm}
\section{Conclusion}
In this paper, we proposed \method for cross-domain speech recognition.
Our key motivation is to match the character-level distributions from the source and target domain to leverage the fine-grained information for better adaptation.
Experiments on cross-device and cross-environment ASR have shown the superiority of our \method.
Moreover, we also conduct extensive experiments to empirically analyze the contribution of encoders and decoders in Speech-Transformer architectures and analyze different frame-level label assignment strategies.
In the future, we plan to extend our research into other speech adaptation applications such as dataset adaptation and speaker adaptation.

\bibliographystyle{IEEEtran}

\bibliography{mybib}

\begin{thebibliography}{10}
\providecommand{\url}[1]{#1}
\csname url@samestyle\endcsname
\providecommand{\newblock}{\relax}
\providecommand{\bibinfo}[2]{#2}
\providecommand{\BIBentrySTDinterwordspacing}{\spaceskip=0pt\relax}
\providecommand{\BIBentryALTinterwordstretchfactor}{4}
\providecommand{\BIBentryALTinterwordspacing}{\spaceskip=\fontdimen2\font plus
\BIBentryALTinterwordstretchfactor\fontdimen3\font minus
  \fontdimen4\font\relax}
\providecommand{\BIBforeignlanguage}[2]{{%
\expandafter\ifx\csname l@#1\endcsname\relax
\typeout{** WARNING: IEEEtran.bst: No hyphenation pattern has been}%
\typeout{** loaded for the language `#1'. Using the pattern for}%
\typeout{** the default language instead.}%
\else
\language=\csname l@#1\endcsname
\fi
#2}}
\providecommand{\BIBdecl}{\relax}
\BIBdecl

\bibitem{pan2009survey}
S.~J. Pan and Q.~Yang, ``A survey on transfer learning,'' \emph{IEEE
  Transactions on knowledge and data engineering}, vol.~22, no.~10, pp.
  1345--1359, 2009.

\bibitem{patel2015visual}
V.~M. Patel, R.~Gopalan, R.~Li, and R.~Chellappa, ``Visual domain adaptation: A
  survey of recent advances,'' \emph{IEEE signal processing magazine}, vol.~32,
  no.~3, pp. 53--69, 2015.

\bibitem{liang2018learning}
D.~Liang, Z.~Huang, and Z.~C. Lipton, ``Learning noise-invariant
  representations for robust speech recognition,'' in \emph{IEEE Spoken
  Language Technology Workshop (SLT)}.\hskip 1em plus 0.5em minus 0.4em\relax
  IEEE, 2018, pp. 56--63.

\bibitem{khurana2020unsupervised}
S.~Khurana, N.~Moritz, T.~Hori, and J.~L. Roux, ``Unsupervised domain
  adaptation for speech recognition via uncertainty driven self-training,''
  \emph{arXiv preprint arXiv:2011.13439}, 2020.

\bibitem{sun2018domain}
S.~Sun, C.-F. Yeh, M.-Y. Hwang, M.~Ostendorf, and L.~Xie, ``Domain adversarial
  training for accented speech recognition,'' in \emph{IEEE International
  Conference on Acoustics, Speech and Signal Processing (ICASSP)}.\hskip 1em
  plus 0.5em minus 0.4em\relax IEEE, 2018, pp. 4854--4858.

\bibitem{duan2020unsupervised}
R.~Duan, ``Unsupervised feature adaptation using adversarial multi-task
  training for automatic evaluation of children’s speech,'' in \emph{ICASSP},
  2020.

\bibitem{ganin2016domain}
Y.~Ganin, E.~Ustinova, H.~Ajakan, P.~Germain, H.~Larochelle, F.~Laviolette,
  M.~Marchand, and V.~Lempitsky, ``Domain-adversarial training of neural
  networks,'' \emph{The journal of machine learning research}, vol.~17, no.~1,
  pp. 2096--2030, 2016.

\bibitem{zhu2020deep}
Y.~Zhu, F.~Zhuang, J.~Wang, G.~Ke, J.~Chen, J.~Bian, H.~Xiong, and Q.~He,
  ``Deep subdomain adaptation network for image classification,'' \emph{IEEE
  transactions on neural networks and learning systems}, 2020.

\bibitem{graves2006connectionist}
A.~Graves, S.~Fern{\'a}ndez, F.~Gomez, and J.~Schmidhuber, ``Connectionist
  temporal classification: labelling unsegmented sequence data with recurrent
  neural networks,'' in \emph{Proceedings of the 23rd international conference
  on Machine learning}, 2006, pp. 369--376.

\bibitem{gretton2012kernel}
A.~Gretton, K.~M. Borgwardt, M.~J. Rasch, B.~Sch{\"o}lkopf, and A.~Smola, ``A
  kernel two-sample test,'' \emph{The Journal of Machine Learning Research},
  vol.~13, no.~1, pp. 723--773, 2012.

\bibitem{masumura2020sequence}
R.~Masumura, M.~Ihori, A.~Takashima, T.~Moriya, A.~Ando, and Y.~Shinohara,
  ``Sequence-level consistency training for semi-supervised end-to-end
  automatic speech recognition,'' in \emph{ICASSP 2020-2020 IEEE International
  Conference on Acoustics, Speech and Signal Processing (ICASSP)}.\hskip 1em
  plus 0.5em minus 0.4em\relax IEEE, 2020, pp. 7054--7058.

\bibitem{dong2018speech}
L.~Dong, S.~Xu, and B.~Xu, ``Speech-transformer: a no-recurrence
  sequence-to-sequence model for speech recognition,'' in \emph{IEEE
  International Conference on Acoustics, Speech and Signal Processing
  (ICASSP)}.\hskip 1em plus 0.5em minus 0.4em\relax IEEE, 2018, pp. 5884--5888.

\bibitem{mathur2020libri}
A.~Mathur, F.~Kawsar, N.~Berthouze, and N.~D. Lane, ``Libri-adapt: a new speech
  dataset for unsupervised domain adaptation,'' in \emph{ICASSP 2020-2020 IEEE
  International Conference on Acoustics, Speech and Signal Processing
  (ICASSP)}.\hskip 1em plus 0.5em minus 0.4em\relax IEEE, 2020, pp. 7439--7443.

\bibitem{kim2017joint}
S.~Kim, T.~Hori, and S.~Watanabe, ``Joint ctc-attention based end-to-end speech
  recognition using multi-task learning,'' in \emph{2017 IEEE international
  conference on acoustics, speech and signal processing (ICASSP)}.\hskip 1em
  plus 0.5em minus 0.4em\relax IEEE, 2017, pp. 4835--4839.

\bibitem{karita2019comparative}
S.~Karita, N.~Chen, T.~Hayashi, T.~Hori, H.~Inaguma, Z.~Jiang, M.~Someki,
  N.~E.~Y. Soplin, R.~Yamamoto, X.~Wang \emph{et~al.}, ``A comparative study on
  transformer vs rnn in speech applications,'' in \emph{2019 IEEE Automatic
  Speech Recognition and Understanding Workshop (ASRU)}.\hskip 1em plus 0.5em
  minus 0.4em\relax IEEE, 2019, pp. 449--456.

\bibitem{Hou2020}
\BIBentryALTinterwordspacing
W.~Hou, Y.~Dong, B.~Zhuang, L.~Yang, J.~Shi, and T.~Shinozaki, ``{Large-Scale
  End-to-End Multilingual Speech Recognition and Language Identification with
  Multi-Task Learning},'' in \emph{Proc. Interspeech 2020}, 2020, pp.
  1037--1041. [Online]. Available:
  \url{http://dx.doi.org/10.21437/Interspeech.2020-2164}
\BIBentrySTDinterwordspacing

\bibitem{hou2021meta}
W.~Hou, Y.~Wang, S.~Gao, and T.~Shinozaki, ``Meta-adapter: Efficient
  cross-lingual adaptation with meta-learning,'' in \emph{ICASSP 2021-2021 IEEE
  International Conference on Acoustics, Speech and Signal Processing
  (ICASSP)}.\hskip 1em plus 0.5em minus 0.4em\relax IEEE, 2021, pp. 7028--7032.

\bibitem{hou2021exploiting}
W.~Hou, H.~Zhu, Y.~Wang, J.~Wang, T.~Qin, R.~Xu, and T.~Shinozaki, ``Exploiting
  adapters for cross-lingual low-resource speech recognition,'' \emph{arXiv
  preprint arXiv:2105.11905}, 2021.

\bibitem{vaswani2017attention}
A.~Vaswani, N.~Shazeer, N.~Parmar, J.~Uszkoreit, L.~Jones, A.~N. Gomez,
  {\L}.~Kaiser, and I.~Polosukhin, ``Attention is all you need,''
  \emph{Advances in Neural Information Processing Systems}, vol.~30, pp.
  5998--6008, 2017.

\bibitem{pan2019transferrable}
Y.~Pan, T.~Yao, Y.~Li, Y.~Wang, C.-W. Ngo, and T.~Mei, ``Transferrable
  prototypical networks for unsupervised domain adaptation,'' in
  \emph{Proceedings of the IEEE/CVF Conference on Computer Vision and Pattern
  Recognition}, 2019, pp. 2239--2247.

\bibitem{sak2015learning}
H.~Sak, A.~Senior, K.~Rao, O.~Irsoy, A.~Graves, F.~Beaufays, and J.~Schalkwyk,
  ``Learning acoustic frame labeling for speech recognition with recurrent
  neural networks,'' in \emph{2015 IEEE international conference on acoustics,
  speech and signal processing (ICASSP)}.\hskip 1em plus 0.5em minus
  0.4em\relax IEEE, 2015, pp. 4280--4284.

\bibitem{kurzinger2020ctc}
L.~K{\"u}rzinger, D.~Winkelbauer, L.~Li, T.~Watzel, and G.~Rigoll,
  ``Ctc-segmentation of large corpora for german end-to-end speech
  recognition,'' in \emph{International Conference on Speech and
  Computer}.\hskip 1em plus 0.5em minus 0.4em\relax Springer, 2020, pp.
  267--278.

\bibitem{borgwardt2006integrating}
K.~M. Borgwardt, A.~Gretton, M.~J. Rasch, H.-P. Kriegel, B.~Sch{\"o}lkopf, and
  A.~J. Smola, ``Integrating structured biological data by kernel maximum mean
  discrepancy,'' \emph{Bioinformatics}, vol.~22, no.~14, pp. e49--e57, 2006.

\bibitem{pan2010domain}
S.~J. Pan, I.~W. Tsang, J.~T. Kwok, and Q.~Yang, ``Domain adaptation via
  transfer component analysis,'' \emph{IEEE Transactions on Neural Networks},
  vol.~22, no.~2, pp. 199--210, 2010.

\bibitem{tzeng2014deep}
E.~Tzeng, J.~Hoffman, N.~Zhang, K.~Saenko, and T.~Darrell, ``Deep domain
  confusion: Maximizing for domain invariance,'' \emph{arXiv preprint
  arXiv:1412.3474}, 2014.

\bibitem{zhang2017joint}
J.~Zhang, W.~Li, and P.~Ogunbona, ``Joint geometrical and statistical alignment
  for visual domain adaptation,'' in \emph{Proceedings of the IEEE conference
  on computer vision and pattern recognition}, 2017, pp. 1859--1867.

\bibitem{shivakumar2020transfer}
P.~G. Shivakumar and P.~Georgiou, ``Transfer learning from adult to children
  for speech recognition: Evaluation, analysis and recommendations,''
  \emph{Computer speech \& language}, vol.~63, p. 101077, 2020.

\bibitem{Watanabe2018}
S.~Watanabe, T.~Hori, S.~Karita, T.~Hayashi, J.~Nishitoba, Y.~Unno, N.~{Enrique
  Yalta Soplin}, J.~Heymann, M.~Wiesner, N.~Chen, A.~Renduchintala, and
  T.~Ochiai, ``Espnet: End-to-end speech processing toolkit,'' in \emph{Proc.
  Interspeech}, 2018.

\end{thebibliography}

\end{document}